\documentclass[final,5p,times,twocolumn]{elsarticle}

\usepackage{amssymb}
\usepackage{amsmath}
\usepackage{xcolor}
\usepackage[normalem]{ulem} 

\begin{document}
	
	\begin{frontmatter}
		
		
		
		\title{Emergence of solitary and chimera states in adaptive pendulum networks under diverse learning rules}
		
		
		\author[inst1]{R. Anand}
		
		\affiliation[inst1]{organization={Centre for Nonlinear Science and Engineering, Department of Physics, SASTRA Deemed to be University},
			addressline={}, 
			city={Thanjavur},
			postcode={613401}, 
			state={TamilNadu},
			country={India}}
			
		\author[inst1]{V. K. Chandrasekar}
		\author[inst1]{R. Suresh}
		\ead{suresh@eee.sastra.edu}
		
		\begin{abstract}
We investigate the interplay between phase lag and adaptive learning rules in a network of identical pendulum oscillators, where the coupling strengths evolve dynamically in response to the oscillators’ states. Specifically, we examine two biologically inspired adaptation mechanisms, Hebbian and spike-timing-dependent plasticity (STDP), and their influence on the emergence of collective dynamical patterns. Under Hebbian adaptation, the network exhibits a wide range of organized behaviors, including two-cluster, solitary, multi-antipodal, and chimera states. In contrast, STDP coupling induces splay, splay-cluster, and splay-chimera configurations. Importantly, we find that the solitary state arises spontaneously in this adaptive network without requiring delays, nonlocal coupling, or external perturbations; instead, it is induced purely by variations in the phase-lag parameter. To the best of our knowledge, such delay-free and symmetry-preserving emergence of solitary behavior has not been reported previously in adaptive oscillator systems. To systematically characterize the resulting dynamical transitions, we employ two complementary incoherence measures based on the local standard deviation of (i) time-averaged frequencies and (ii) instantaneous phases across spatial bins, enabling the construction of detailed two-parameter phase diagrams. Analytical stability analysis of the two-cluster state shows strong agreement with numerical simulations, revealing regions of pronounced multistability. These findings establish adaptive pendulum networks as a minimal yet powerful framework for studying self-organized synchronization, chimera formation, and multistable transitions driven by diverse adaptation mechanisms.
		\end{abstract}
		
	\end{frontmatter}
	
\section{Introduction} \label{intro}
The study of complex networks provides a fundamental framework for understanding collective dynamics observed across a variety of natural and engineered systems, from neuronal assemblies to power grids and social structures \cite{boccaletti2006complex,arenas2008synchronization}. Research on network dynamics has traditionally focused on the emergence of synchronization, wherein individual oscillators evolve coherently \cite{grinstein2005synchronous}. However, subsequent studies, particularly in neural networks subject to stochastic noise \cite{brunel1999fast}, have revealed that large-scale oscillatory behavior can spontaneously emerge even when individual elements remain desynchronized. Such emergent collective dynamics have been explored under diverse conditions, including time delays, heterogeneity among oscillators, and time-dependent coupling strengths \cite{timme2002coexistence, denker2004breaking, maistrenko2007multistability}.

Within this context, adaptive networks, systems in which the connection topology evolves concurrently with the nodal dynamics, have attracted substantial attention \cite{taylor2010spontaneous,horstmeyer2020adaptive, markram1997regulation, abbott2000synaptic, jain2001model,kuehn2015multiple}. Adaptation mechanisms enable feedback between network structure and dynamics, leading to self-organized synchronization and other rich dynamic phenomena. For example, slow adaptation relative to the intrinsic timescale of the oscillators allows the use of singular perturbation analysis, revealing how long-term plasticity shapes collective states \cite{kuehn2015multiple}.

Adaptive dynamical networks are known to exhibit a variety of self-organized patterns, including multi-cluster synchronization, chimera states, and hierarchical frequency groupings, arising from the interplay between state evolution and topological adaptation  \cite{seliger2002plasticity, ren2007adaptive,aoki2009co, aoki2011self,abbott2000synaptic}. Gradual adaptation is particularly crucial for stabilizing multi-cluster configurations, where transient formations such as double-antipodal clusters strongly influence long-term dynamics \cite{berner2019hierarchical}. These networks can sustain partially synchronized or hierarchically organized states, such as splay, antipodal, and double-antipodal patterns, through a delicate balance between coupling evolution and network structure \cite{berner2020birth}. Similar mechanisms have been observed in adaptive systems ranging from Morris--Lecar neuronal models \cite{popovych2015spacing} to phase oscillators with dynamic couplings  \cite{kasatkin2017self, kasatkin2018synchronization, kasatkin2018effect, berner2019hierarchical}.

Recent investigations have further revealed heterogeneous nucleation phenomena  \cite{fialkowski2023heterogeneous,yadav2024disparity} and frequency-dependent synchronization shifts in multilayer adaptive networks \cite{yadav2025heterogeneous}. Such systems often feature slowly evolving coupling variables, which give rise to unique dynamical phenomena, including Canard cascades \cite{balzer2024canard}.

The formulation of adaptation rules is largely inspired by neuronal learning principles, particularly those describing the pre- and postsynaptic timing of neuronal firing \cite{gerstner1998neuronal, caporale2008spike, bi2001synaptic, meisel2009adaptive, lucken2016noise}. Among these, Hebbian and spike-timing-dependent plasticity (STDP) rules remain the most extensively studied. Hebb’s postulate, summarized as “neurons that fire together wire together” \cite{donald1949organization,lowel1992selection,zhang1998critical}--implies that coupling strengths increase for in-phase firing and weaken for antiphase activity. These principles form the foundation for long-term memory formation and neuromorphic computing architectures \cite{acebron2005kuramoto,pickett2013scalable,hansen2017double,birkoben2020slow}. Understanding how learning rules interact with phase delays is therefore essential for describing the emergent behaviours of adaptive systems \cite{timms2014synchronization, thamizharasan2025dynamics}.

Prior studies have reported a wide range of collective patterns, including stable phase clusters, antipodal clusters, chimera states, and explosive synchronization, in adaptive or multiplexed oscillator networks \cite{berner2020birth,berner2019multiclusters,zhang2015explosive, kumar2020interlayer}. However, these works primarily focused on first-order phase oscillators or networks where solitary states appear only under nonlocal coupling or delay-induced effects  \cite{berner2020solitary, thamizharasan2025dynamics}.

In contrast, the present work investigates, for the first time, the collective dynamics of identical pendulum oscillators governed by adaptive coupling with an intrinsic phase lag, in the absence of any restoring force. We demonstrate that the system supports a wide spectrum of novel dynamical states: two-cluster (TC), solitary (SS), multi-antipodal (MAC), chimera (CHI), splay (SP), splay-cluster (SPC), splay-chimera (SPCHI), and incoherent (INC) states depending on the phase lag and adaptation rule. Importantly, we find that solitary states emerge in this globally coupled adaptive network without requiring delays, nonlocal coupling, or external forcing; instead, they arise purely from variations in the phase-lag parameter.

To systematically characterize these dynamical behaviors, we employ two independent incoherence measures based on the local standard deviation of (i) time-averaged frequencies and (ii) instantaneous phases across spatial bins. These measures enable the construction of detailed two-parameter phase diagrams that clearly delineate the transitions between the observed dynamical regimes. We also provide an analytical stability analysis of the two-cluster state, showing strong agreement with numerical results and highlighting regions of multistability within the adaptive pendulum network.
\begin{figure}[!ht]
	\centering
	\includegraphics[width=0.5\textwidth]{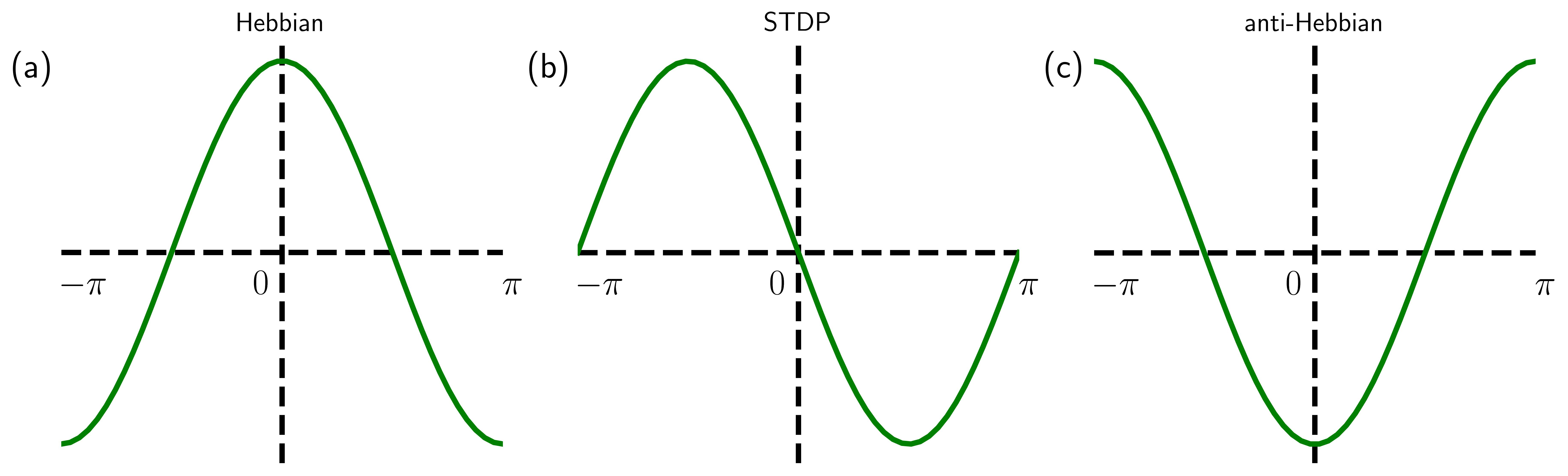}
	\caption{Evolution of coupling strengths under different adaptation rules characterized by the control parameter $\beta$. (a) Hebbian adaptation for $\beta = -\pi/2$, (b) STDP for $\beta = 0$, and (c) anti-Hebbian adaptation for $\beta = \pi/2$. }
	\label{hebbian}
\end{figure}

The remainder of this paper is organized as follows: Section \ref{model} presents the model formulation of the adaptive pendulum network. Section \ref{dynamics} discusses the dynamical states observed under different adaptation rules.
In Sec. \ref{measure}, we analyze dynamical transitions and characterization measures using one- and two-parameter phase diagrams. Finally, Sec. \ref{conclusion} summarizes the main findings and their implications.

\section{Model}\label{model}
A key extension of the Kuramoto model to pendulum networks involves incorporating a second-order inertial term to capture the oscillators’ rotational dynamics  \cite{ebrahimzadeh2022mixed, li2023chimera}. This modification was originally developed within an adaptive framework to explain the precise synchronization of flashing in Pteroptyx malaccae fireflies, which achieve near-zero phase differences through slow adaptation \cite{taylor2010spontaneous}. Similar adaptive mechanisms have since been applied to model diverse biological, chemical, and technological systems, including power grid dynamics \cite{sawicki2022modeling, berner2022critical, berner2021adaptive}.

In the present study, we consider a globally coupled network of $N$ identical pendulum oscillators, where each node interacts with every other node through adaptive coupling. The dynamics of the $i^{\mathrm{th}}$ oscillator are governed by the evolution equations
\begin{eqnarray}
	\label{sys2}
	\ddot{\phi}_i + \dot{\phi}_i &=& \nu - \frac{1}{N}\sum_{j=1}^{N}{k}_{ij}\sin(\phi_i - \phi_j + \alpha), \\
	\dot{k}_{ij} &=& -\epsilon (\sin(\phi_i - \phi_j + \beta) + {k}_{ij}), \; \nonumber 
\end{eqnarray}

where $\phi_i$ denotes the phase, $\dot{\phi}_i$ the angular velocity, and $\ddot{\phi}_i$ the angular acceleration of the $i^{\mathrm{th}}$ pendulum. The parameter $\nu$ represents the intrinsic frequency, and $k_{ij}$ is the coupling strength from node $j$ to node $i$, bounded by $|k_{ij}| \leq 1$.

The small positive parameter $\epsilon \ll 1$ determines the timescale separation between the fast nodal dynamics and the slow adaptation of the coupling weights. The phase-lag parameter $\alpha$ introduces a shift in the interaction term, while $\beta$ controls the adaptation rule, determining whether the system follows STDP, Hebbian, or anti-Hebbian dynamics \cite{aoki2009co, aoki2011self}.

As illustrated schematically in Fig. \ref{hebbian}, $\beta = 0$ corresponds to STDP, $\beta = -\pi/2$ corresponds to the Hebbian rule, and  $\beta = \pi/2$ corresponds to the anti-Hebbian rule. To ensure bounded coupling, we impose the saturation limits $k_{ij}=1$ if $k_{ij}>1$ and $k_{ij}=-1$ if $k_{ij}<-1$. In this study, we impose periodic boundary conditions on the oscillator phases, defined on the interval $[0,2\pi)$.

\begin{figure}[!ht]
	\centering
	\includegraphics[width=1.0\linewidth]{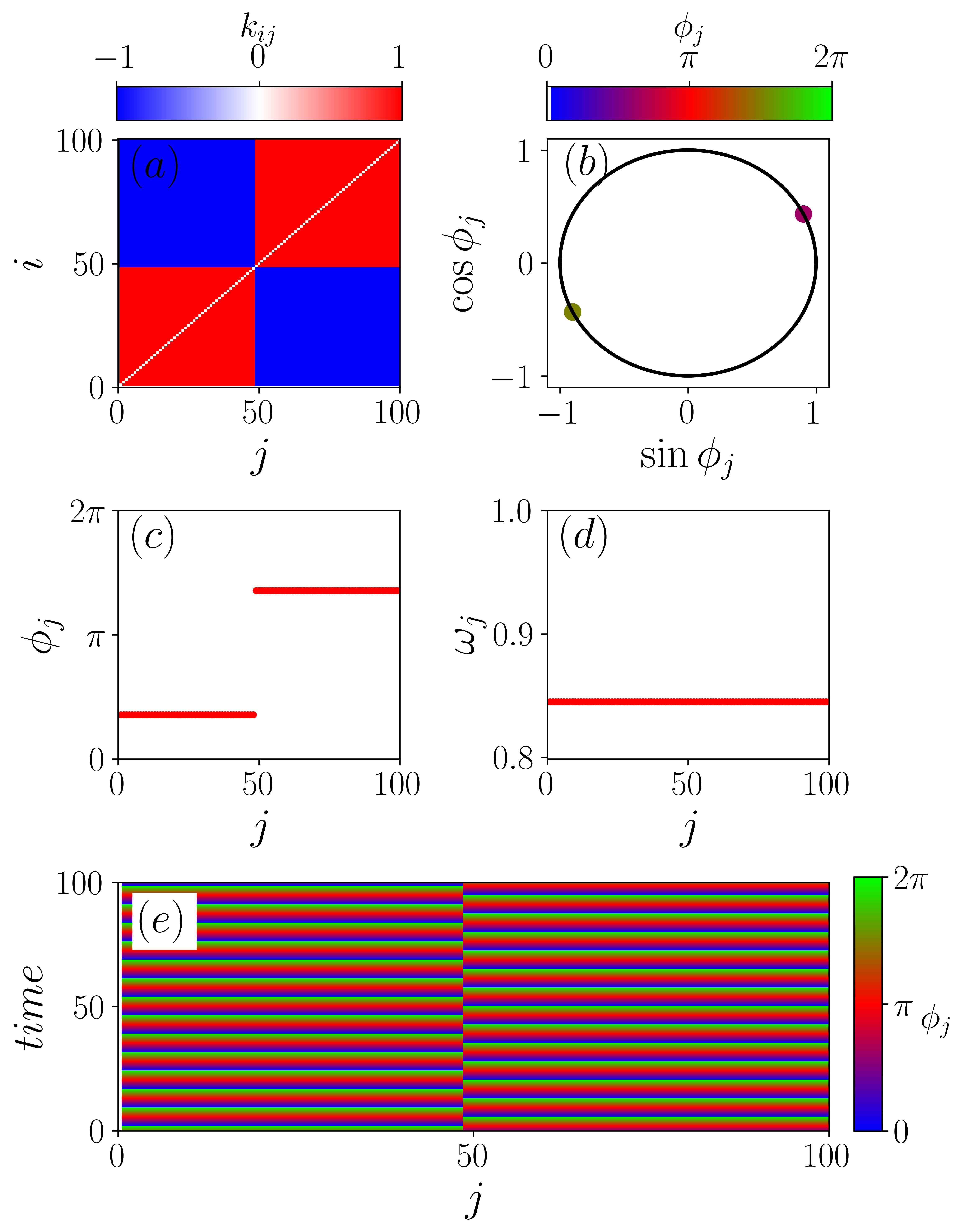}
	\caption{(a) Coupling matrix $k_{ij}$ corresponding to the two cluster state. (b) Phase portrait of the instantaneous phases. (c) Snapshot of the instantaneous phases and (d) averaged frequency profile. (e) Space-time evolution of the TC state for $\alpha = 0.05\pi$. The other parameters are fixed as $\nu = 1.0$, $\epsilon = 0.01$, and $\beta = -0.5\pi$.}
	\label{Fig1}
\end{figure}
Although these adaptation rules originate from neuronal models, the phase variables $\phi_i$ can be interpreted as analogues of spike timing in weakly coupled oscillators near a Hopf bifurcation \cite{seliger2002young, aoyagi1997effect}. Consequently, the pendulum variables, being time-averaged representations of experimentally observable oscillations, allow the system to reproduce neuronal-like synchronization phenomena \cite{schuster1990model}.

In earlier studies, similar adaptation mechanisms have produced diverse dynamical states in identical phase-oscillator systems subject to external stimuli and bounded Hebbian/anti-Hebbian couplings \cite{thamizharasan2022exotic, thamizharasan2024hebbian, thamizharasan2024stimulus}. In the present model, we uncover eight distinct collective states arising from Eq. (\ref{sys2}): two-cluster (TC), solitary state (SS), multi-antipodal cluster state (MAC), chimera (CHI), splay (SP), splay cluster (SPC), and splay chimera (SPCHI) states.

To identify and distinguish these states, we analyze the oscillators’ instantaneous phases, angular frequencies, and coupling matrices. The average frequency of each oscillator is computed after a sufficiently long transient, ensuring that all temporal fluctuations have decayed. In the present simulations, the transient interval is set to $t \in (27400, 27500)$, providing statistically steady dynamical regimes.
\begin{figure}[!ht]
	\centering
	\includegraphics[width=1.0\linewidth]{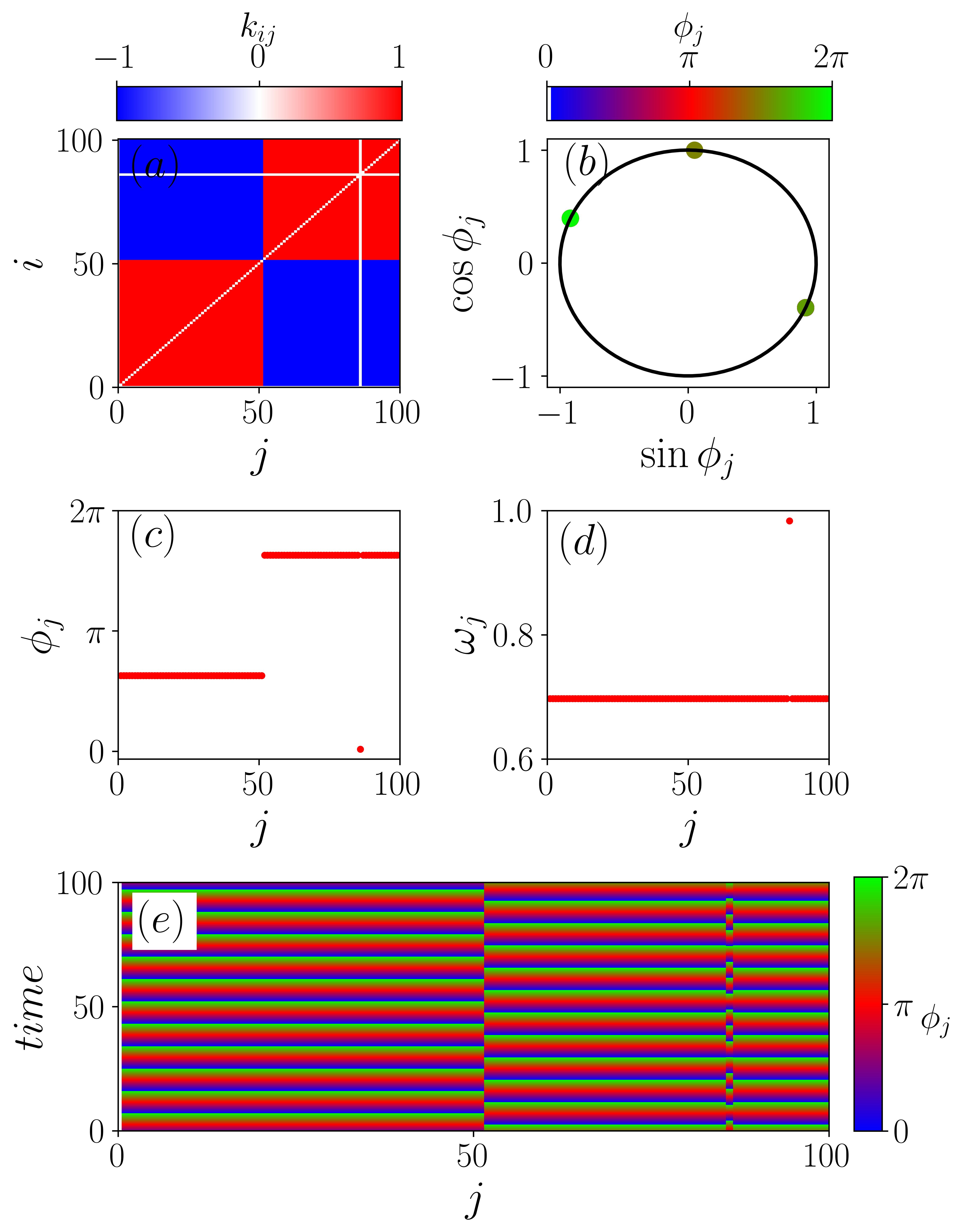}
	\caption{(a) Coupling matrix $k_{ij}$ corresponding to the solitary state. (b) Phase portrait of the instantaneous phases. (c) Snapshot of the instantaneous phases and (d) averaged frequency profile. (e) Space-time evolution of the SS state for $\alpha = 0.12\pi$.}
	\label{Fig2}
\end{figure}
\section{Dynamical States}\label{dynamics}	
The coupled adaptive pendulum network described by Eq. (\ref{sys2}) is numerically integrated using a fourth-order Runge-Kutta algorithm with a fixed time step of 0.01. The network size is fixed at $N = 100$, and unless otherwise specified, the parameters are set as $\nu = 1$ and $\epsilon = 0.01$. The initial phases $\phi_i$ are randomly distributed in the interval $[0, 2\pi)$, the initial velocities $\dot{\phi}_i$ are uniformly chosen from $(-0.5, 0.5)$, and the coupling weights $k_{ij}$ are selected from $(-1, 1)$. The network dynamics strongly depend on the phase-lag parameter $\alpha$ and the adaptation parameter $\beta$, resulting in a variety of collective states. These states are identified by analyzing the coupling matrices, instantaneous phase portraits, phase snapshots, average frequency distributions, and space-time plots of the oscillators.

\textbf{A. Two-Cluster (TC) State:}  
For small phase-lag values ($\alpha = 0.05\pi$) under the Hebbian adaptation rule ($\beta = -0.5\pi$), the oscillators organize into two synchronized clusters, termed as TC state \cite{aoki2009co, aoki2011self} oscillating in antiphase. Within each cluster, the oscillators maintain perfect phase synchronization, while the two clusters differ by a constant phase shift of $\pi$. Figure \ref{Fig1}(a) shows the coupling matrix $k_{ij}$, where intra-cluster couplings take the value $k_{ij} = 1$ and inter-cluster couplings take $k_{ij} = -1$, indicating excitatory and inhibitory interactions, respectively. The instantaneous phase portrait in Fig. \ref{Fig1}(b) displays two distinct groups separated by $\pi$, confirming the antiphase relation. The corresponding phase snapshot, average frequency, and space-time plots are shown in Figs. \ref{Fig1}(c)-\ref{Fig1}(e), clearly indicate the coexistence of two perfectly synchronized antiphase populations. This configuration represents a symmetric and stable partition of the network, characteristic of the Hebbian coupling rule.
\begin{figure}[!ht]
	\centering
	\includegraphics[width=1.0\linewidth]{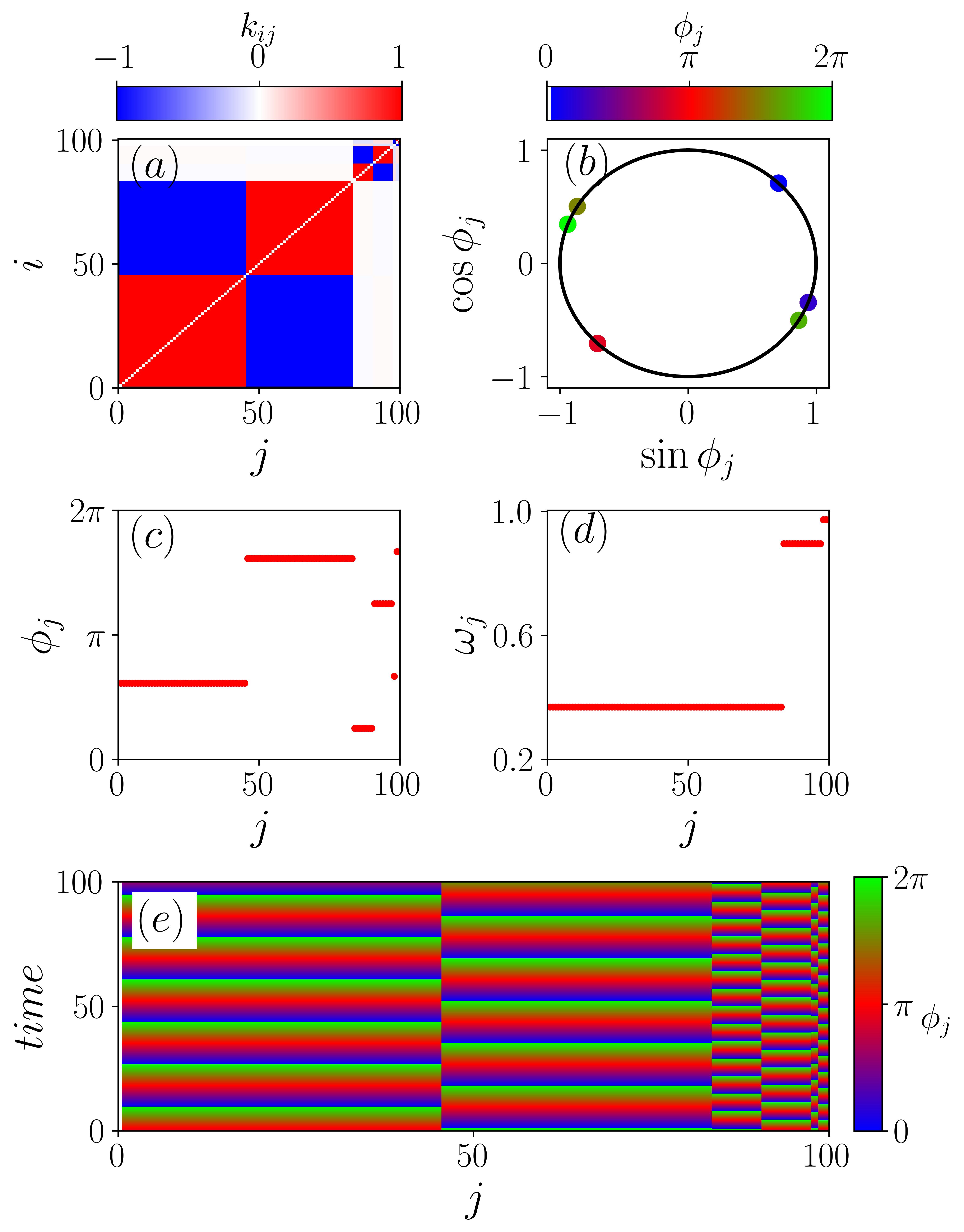}
	\caption{(a) Coupling matrix $k_{ij}$ corresponding to the multi-antipodal cluster state. (b) Phase portrait of the instantaneous phases. (c) Snapshot of the instantaneous phases and (d) averaged frequency profile. (e) Space-time evolution of the MAC state for $\alpha = 0.25\pi$.}
	\label{Fig3}
\end{figure}

\textbf{B. Solitary State (SS):}  
Increasing the phase lag to $\alpha = 0.12\pi$ induces a transition from the TC state to a solitary state. In this configuration, most oscillators remain frequency synchronized, while one or a few oscillators detach from the synchronized group and oscillate with distinct frequencies\cite{ebrahimzadeh2022mixed, vikram2025emergence, jaros2018solitary}. The coupling matrix in Fig. \ref{Fig2}(a) reflects localized variations in the interaction strength corresponding to the solitary elements. The instantaneous phase portrait in Fig. \ref{Fig2}(b) shows the separation between the synchronized population and the solitary oscillator. The phase snapshot and average frequency plots in Figs. \ref{Fig2}(c) and \ref{Fig2}(d) reveal that all oscillators except one share the same frequency, while the solitary oscillator drifts independently. The space-time evolution in Fig. \ref{Fig2}(e) further demonstrates the persistent deviation of the solitary oscillator from the synchronized manifold. The solitary state emerges spontaneously without any external forcing or delay, arising purely due to the increase in the phase lag parameter.

\begin{figure}[!ht]
	\centering
	\includegraphics[width=1.0\linewidth]{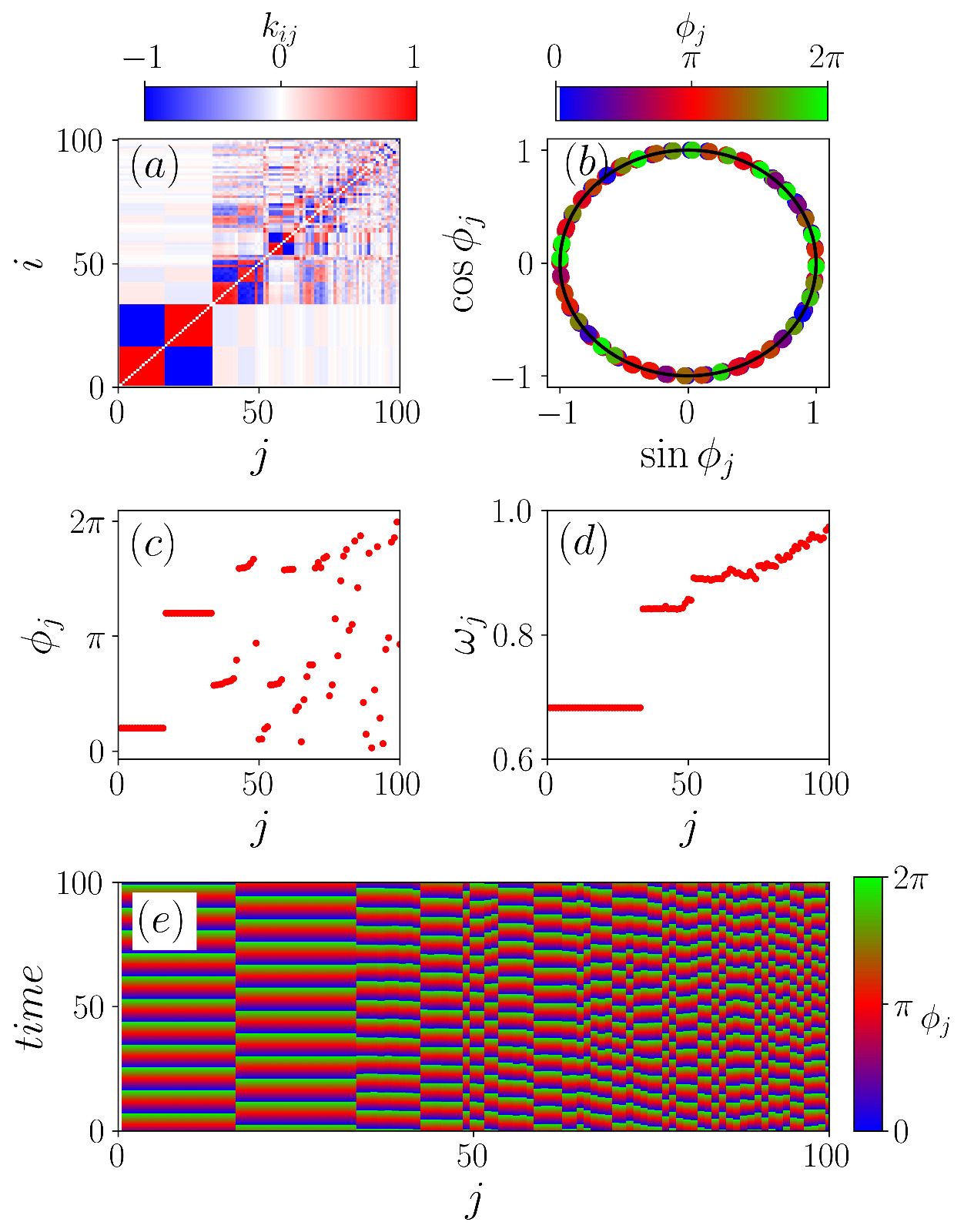}
	\caption{(a) Coupling matrix $k_{ij}$ corresponding to the chimera  state. (b) Phase portrait of the instantaneous phases. (c) Snapshot of the instantaneous phases and (d) averaged frequency profile. (e) Space-time evolution of the CHI state for $\alpha = 0.46\pi$.}
	\label{Fig4}
\end{figure}
\textbf{C. Multi-Antipodal Cluster (MAC) State:}  
As the phase lag increases further to $\alpha = 0.25\pi$, the network evolves into a MAC state. In this state, the oscillators split into three or more clusters oscillating either in phase or in antiphase with each other. The coupling matrix shown in Fig. \ref{Fig3}(a) displays alternating positive and negative blocks corresponding to multiple antipodal groups. The instantaneous phase distribution in Fig. \ref{Fig3}(c) reveals the existence of several evenly spaced phase clusters, and the space-time plot in Fig. \ref{Fig3}(e) confirms their stability over time. The phase portrait and frequency profile in Figs. \ref{Fig3}(b) and \ref{Fig3}(d) show distinct, frequency-locked subpopulations. This configuration satisfies the antipodal relation $2\phi_{ij} = 2\phi_i \pmod{2\pi}$, which generalizes the two-cluster arrangement to higher-order symmetry.

\begin{figure}[!ht]
	\centering
	\includegraphics[width=1.0\linewidth]{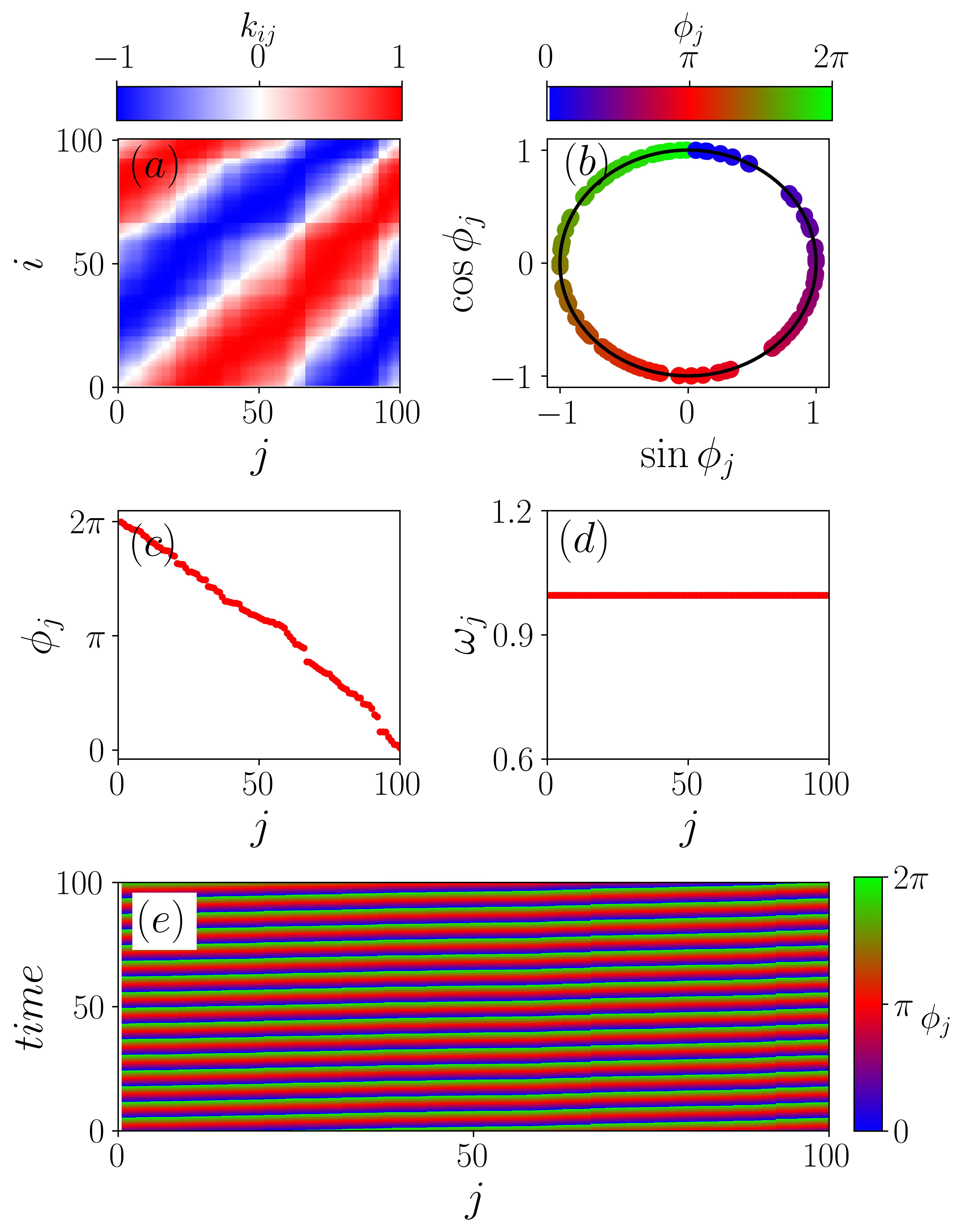}
	\caption{(a) Coupling matrix $k_{ij}$ corresponding to the splay (SP) state. (b) Phase portrait of the instantaneous phases. (c) Snapshot of the instantaneous phases and (d) averaged frequency profile. (e) Space-time evolution of the SP state for $\alpha = 0.25\pi$.}
	\label{Fig5}
\end{figure}
\textbf{D. Chimera (CHI) State:} 
For larger values of the phase-lag parameter ($\alpha = 0.46\pi$), the system evolves into a CHI state, characterized by the coexistence of synchronized and desynchronized subpopulations of identical oscillators \cite{perc2021chimeras, majhi2019chimera}. The corresponding phase and averaged frequency shown in Figs.~\ref{Fig4}(c) and \ref{Fig4}(d)  clearly exhibits a chimera-like structure, reflecting the simultaneous presence of coherent and incoherent dynamical behavior \cite{kasatkin2017self, thamizharasan2024stimulus}. Figure \ref{Fig4}(a) represents the coupling matrix $k_{ij}$. The instantaneous phase portrait in Fig. \ref{Fig4}(b) reveals coherent clusters coexisting with scattered incoherent oscillators. The corresponding space-time plot in Fig. \ref{Fig4}(e) also confirms the coexistence of coherent and incoherent domains. The chimera state observed in this system differs from conventional chimera states, as it emerges in a globally coupled adaptive network and exhibits spatiotemporal intermittency induced by Hebbian adaptation. The sequence of transitions--TC $\rightarrow$ SS $\rightarrow$ MAC $\rightarrow$ CHI -- demonstrates that increasing the phase lag under the Hebbian rule systematically generates progressively complex partially synchronized states.

\begin{figure}[!ht]
	\centering
	\includegraphics[width=1.0\linewidth]{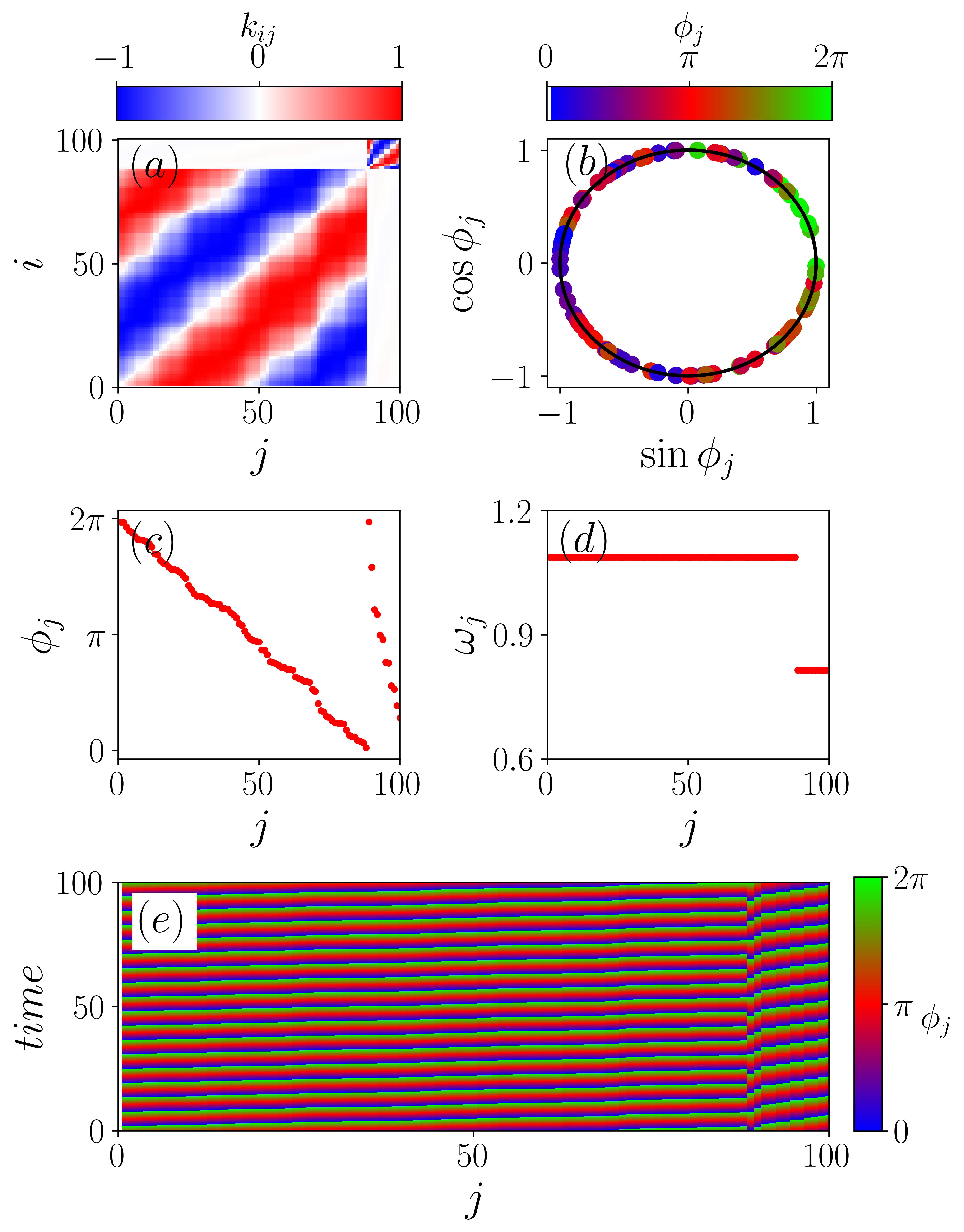}
	\caption{(a) Coupling matrix $k_{ij}$ corresponding to the splay-cluster state. (b) Phase portrait of the instantaneous phases. (c) Snapshot of the instantaneous phases and (d) averaged frequency profile. (e) Space-time evolution of the SPC state for $\alpha = 0.1\pi$.}
	\label{Fig6}
\end{figure}
\textbf{E. Splay (SP) State:} 
Under the STDP rule with $\beta = 0$, the system attains a SP state when the phase-lag parameter is set to $\alpha = 0.25\pi$. In this state, the oscillator phases are uniformly distributed over the unit circle, such that neighboring oscillators are separated by a constant phase difference of $2\pi/N$. Mathematically, the splay state corresponds to a uniformly rotating solution of the form $\phi_i(t) = \Omega t + \psi_i$, where the fixed phase offsets satisfy $\psi_i - \psi_{i-1} = 2\pi/N$ (mod $2\pi$). As a result, all oscillators evolve with an identical instantaneous frequency, $\dot{\phi}_i = \Omega$, leading to a uniform time-averaged frequency across the network, $\omega_i = \Omega$.
	
This dynamical regime is characterized by complete global phase desynchronization, as indicated by a vanishing global order parameter, while simultaneously maintaining strong local phase ordering due to the uniform and constant phase gradient along the oscillator chain. The coupling matrix shown in Fig.~\ref{Fig5}(a) and the space--time plot in Fig.~\ref{Fig5}(e) clearly illustrate this steady linear phase progression across oscillator indices, whereas an instantaneous phase snapshot of the splay state is presented in Fig.~\ref{Fig5}(c). The phase portrait and the time-averaged frequency distributions displayed in Figs.~\ref{Fig5}(b)--\ref{Fig5}(d) further corroborate the regular, uniformly rotating dynamics with identical mean frequencies for all oscillators.

\begin{figure}[!ht]
	\centering
	\includegraphics[width=1.0\linewidth]{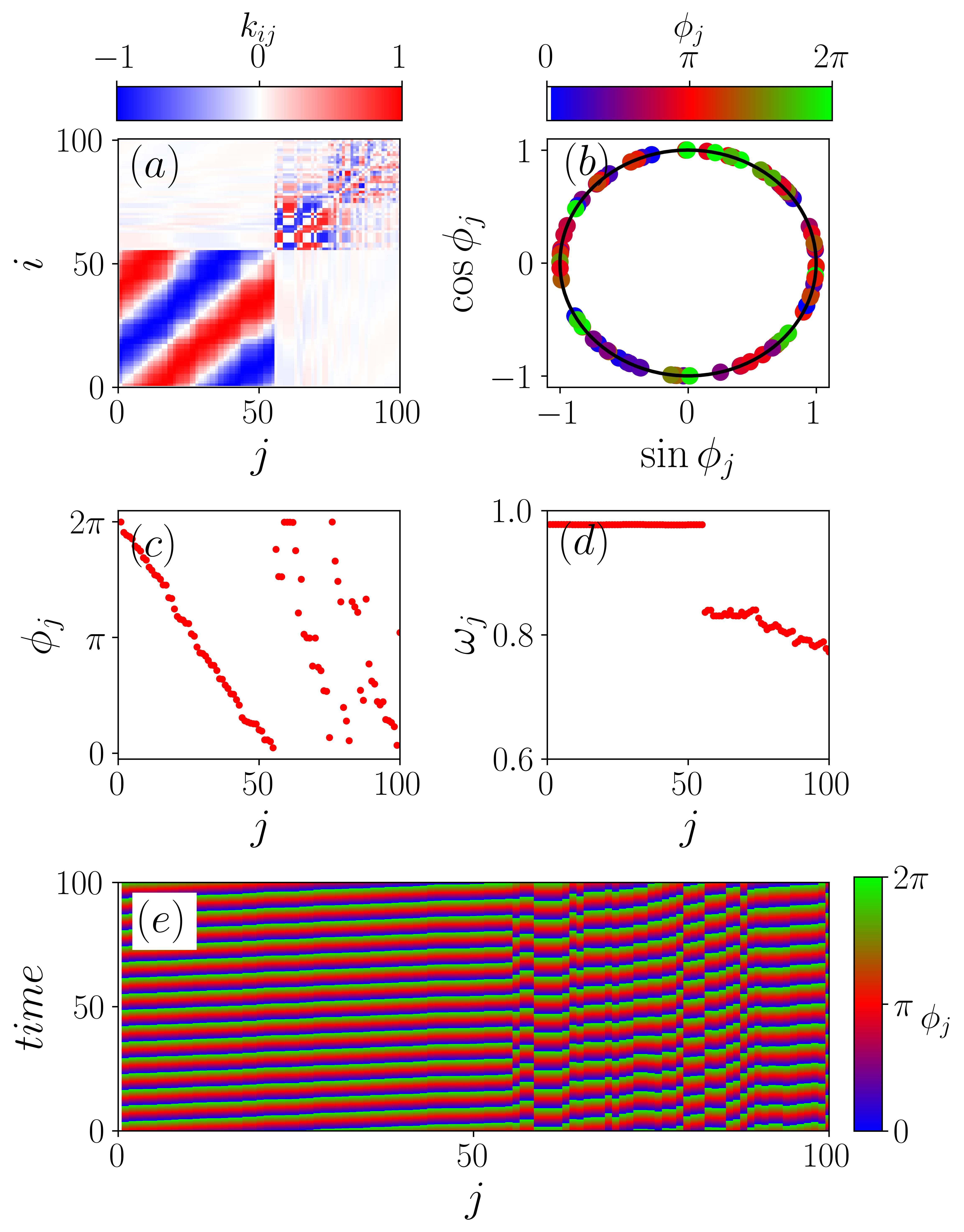}
	\caption{(a) Coupling matrix $k_{ij}$ corresponding to the Splay Chimera State. (b) Phase portrait showing the instantaneous phases. (c) Phase snapshot and (d) corresponding averaged frequency. (e) Space-time plot of the SPCHI state for $\alpha = 0.05\pi$.}
	\label{Fig7}
\end{figure}
\textbf{F. Splay-Cluster (SPC) State:} 
When the phase lag is reduced to $\alpha = 0.1\pi$ under the STDP rule ($\beta$ = 0), the system exhibits a SPC state. In this regime, the oscillators form multiple clusters, each exhibiting an internal splay-like arrangement but with distinct phase offsets between clusters. The coupling matrix in Fig. \ref{Fig6}(a) reveals strong intra-cluster and weaker inter-cluster couplings. This implies that oscillators within the same cluster are more strongly connected to each other, promoting internal phase coherence (splay-like ordering), while the weaker coupling between clusters preventsx their mutual synchronization. Consequently, the clusters maintain distinct phase relationships, giving rise to a partially coherent configuration. The instantaneous phase portrait in Fig. \ref{Fig6}(b) and the phase and frequency distributions in Figs. \ref{Fig6}(c)--\ref{Fig6}(d) display the presence of multiple coherent splay clusters. The space--time evolution shown in Fig. \ref{Fig6}(e) confirms the persistence of these clusters, indicating partial phase coherence and internal ordering. The SPC state thus serves as an intermediate configuration between the perfectly ordered splay and the more complex chimera structures.

\begin{figure}[!ht]
	\centering
	\includegraphics[width=1.0\linewidth]{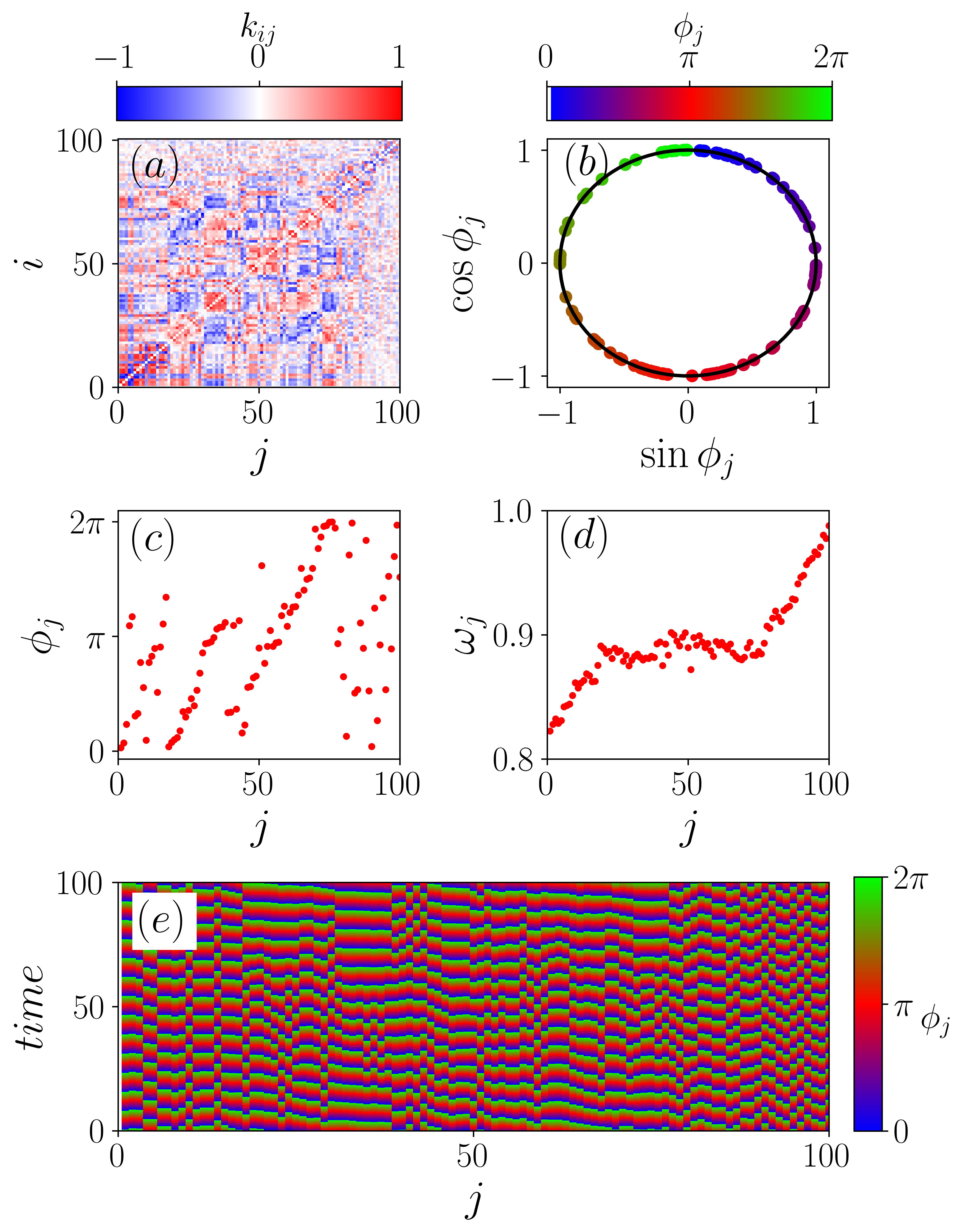}
	\caption{(a) Coupling matrix $k_{ij}$ for the incoherent state. (b) Phase portrait of the instantaneous phases. (c) Phase snapshot and (d) corresponding averaged frequency. (e) Space-time plot of the INC state for $\alpha = 0.4\pi$. }
	\label{Fig8}
\end{figure}
\textbf{G. Splay Chimera (SPCHI) State :} 
For smaller phase-lag values ($\alpha = 0.05\pi$) under STDP adaptation, the network develops a SPCHI state. This hybrid regime combines splay-like order with incoherence, wherein one subpopulation of oscillators maintains a splay configuration, while another subpopulation displays irregular, incoherent dynamics \cite{thamizharasan2024stimulus}. The coupling matrix and phase portraits shown in Figs. \ref{Fig5}(a) and \ref{Fig5}(b) exhibit distinct domains corresponding to the coherent splay group and the incoherent group. The phase snapshots, average frequencies, and space-time plots in Figs. \ref{Fig5}(c)--\ref{Fig5}(e) confirm the coexistence of ordered and disordered behavior. Unlike conventional chimera states, the SPCHI state retains the splay structure within the coherent domain, highlighting the role of adaptive coupling in stabilizing mixed dynamical regimes.

\textbf{H. Incoherent (INC) State :} 
When the adaptation parameter is increased to $\beta = 0.5\pi$ (anti-Hebbian rule), the network loses all coherence and transitions into a completely incoherent state for all values of $\alpha$. In this regime, each oscillator evolves independently, and the phases become randomly distributed without any collective synchronization. The coupling matrix in Fig. \ref{Fig8}(a) shows a disordered pattern, while the instantaneous phase portrait in Fig. \ref{Fig8}(b) and the corresponding phase, frequency, and space-time plots in Figs. \ref{Fig8}(c)--\ref{Fig8}(e) confirm the absence of any coherent structure. The anti-Hebbian adaptation thus destabilizes synchronization and drives the system to a globally incoherent regime.

The adaptive pendulum network exhibits eight distinct collective states: two-cluster, solitary, multi-antipodal cluster, chimera, splay, splay-cluster, splay-chimera, and incoherent, arising from the interplay between the phase lag $\alpha$ and the adaptation rule parameter $\beta$. These states serve as the foundation for constructing the phase diagrams and quantitative measures of synchronization discussed in the next section.

\section{Dynamical Transitions and Characterizations} \label{measure}

\begin{figure}[!ht]
	\centering
	\includegraphics[width=1.0\linewidth]{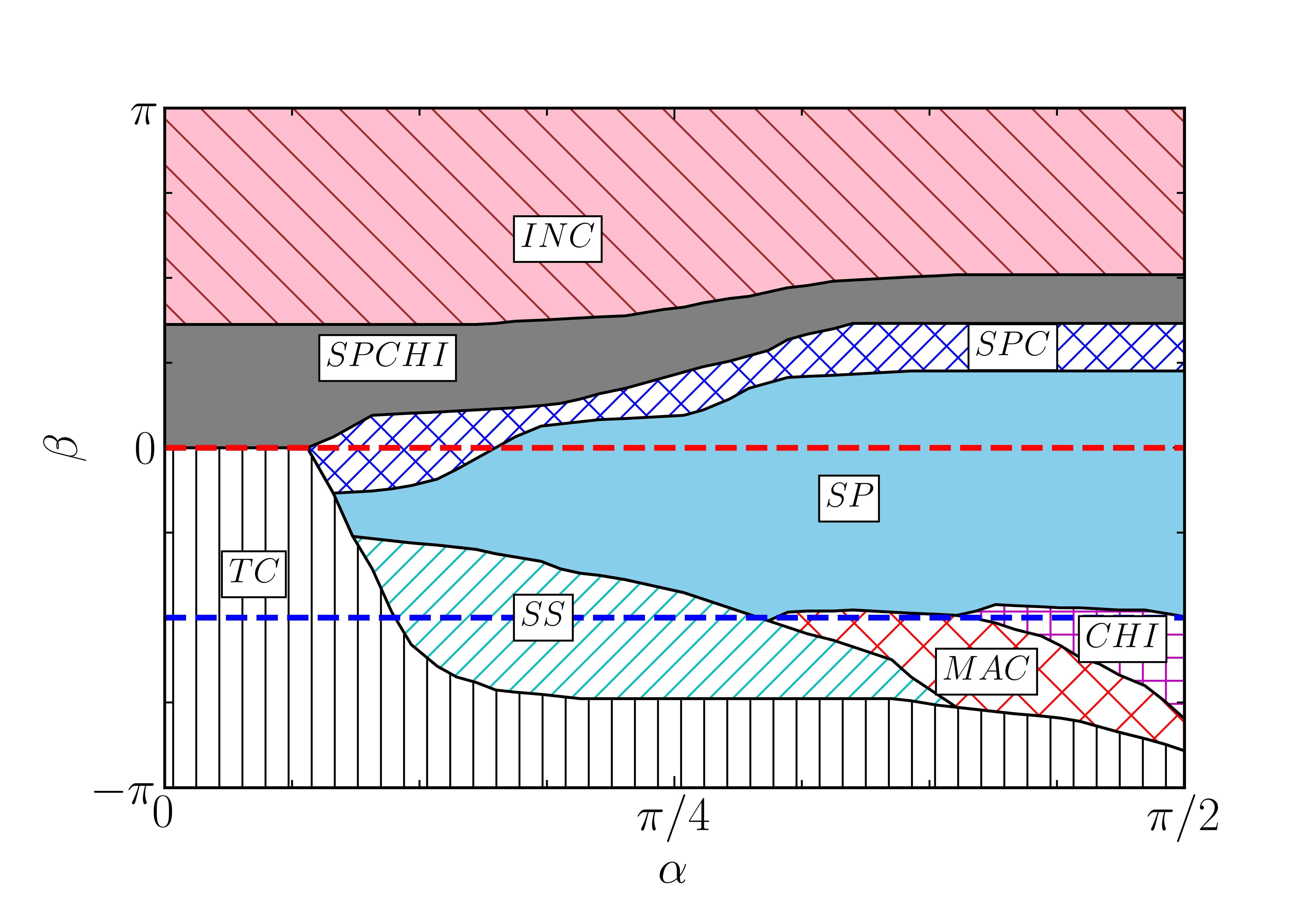}
	\caption{Two-parameter phase diagram of dynamical states in the $(\alpha, \beta)$ space. The initial conditions for $\phi_i$, $\dot{\phi}i$, and $k_{ij}$ are uniformly distributed over the intervals $[0, 2\pi)$, $(-0.5, 0.5)$, and $(-1, 1)$, respectively. The red and blue dashed lines indicate dynamical transitions. The remaining parameters are fixed as $N = 100$, $\nu = 1$, $M = 20$, and $\epsilon = 0.01$.}
	\label{2p}
\end{figure}

In this section, we comprehensively map the diverse dynamical states of the system described by Eq.~(\ref{sys2}) within the two-parameter space defined by $\alpha \in [0, \pi/2)$ and $\beta \in [-\pi, \pi)$, as illustrated in Fig.~\ref{2p}. The boundaries separating these dynamical states in the two-parameter phase diagram are quantitatively characterized using the strength-of-incoherence measures.

To quantify the degree of incoherence among oscillators, we adopt, following Ref. [61], two independent incoherence metrics: the strength of incoherence based on time-averaged frequencies $S$ and its counterpart constructed from the instantaneous phase distribution $S_{\sigma}$. Together, these measures enable a systematic classification of the collective dynamical states and delineate their corresponding regions in the $(\alpha, \beta)$ parameter space. For this purpose, the total number of oscillators $N$ is divided into $M$ bins, each containing $n = N/M$ oscillators.

The local standard deviation of the frequency for each bin, $\sigma_m$, is defined as
\begin{equation}
	\sigma_m = \sqrt{\frac{1}{n}\sum_{j = n(m-1)+1}^{mn} \left[ \omega_j - \bar{\omega}_m \right]^2}, \qquad  m = 1,2,\dots,M,
	\label{eq:sig_m}
\end{equation}
where $\bar{\omega}_m = \frac{1}{n}\sum_{j = n(m-1)+1}^{mn}\omega_j$ is the mean frequency of the $m^{\text{th}}$ bin, and $\omega_j$ denotes the time-averaged frequency of the $j^{\text{th}}$ oscillator.

The \textit{strength of incoherence} $S$ is given by
\begin{equation}
	S = 1 - \frac{1}{M}\sum_{m=1}^{M} s_m, \qquad  s_m = \Theta(\delta - \sigma_m),
	\label{eq:S}
\end{equation}
where $\Theta$ is the Heaviside step function and $\delta = 0.005$ is the fixed threshold used throughout this study.  
For frequency-entrained states such as the two-cluster or splay states, $S = 0$.  
In the solitary state, one oscillator deviates from the synchronized group, producing $\sigma_m > 0$ and $s_m = 0$ for its bin, while $\sigma_m = 0$ and $s_m = 1$ for all others, yielding $S \approx 0$.  
For frequency-clustered states, a discontinuous jump in the mean frequency appears between bins that contain oscillators from distinct clusters, resulting in $s_m = 0$ where $\sigma_m > \delta$ and $s_m = 1$ elsewhere; thus $S = N_{FC}/M \neq 0$, where $N_{FC}$ denotes the number of frequency clusters.  
For chimera and splay-chimera states, $0 < S < 1$, whereas for solitary, multi-antipodal, and splay-cluster states $0 < S \ll 1$.  
In the incoherent state, $S = 1$.

The local standard deviation of the instantaneous phases, $\hat{\sigma}_m$, is defined as
\begin{equation}
	\hat{\sigma}_m = \sqrt{\frac{1}{n}\sum_{j = n(m-1)+1}^{mn}\left[ \phi_j - \bar{\phi}_m \right]^2}, \qquad m = 1,2,\dots,M,
	\label{eq:smb}
\end{equation}
where $\bar{\phi}_m = \frac{1}{n}\sum_{j = n(m-1)+1}^{mn}\phi_j$ is the mean instantaneous phase of the $m^{\text{th}}$ bin, and $\phi_j$ denotes the instantaneous phase of the $j^{\text{th}}$ oscillator.  
This measure helps identify local phase clustering.

The \textit{phase-based strength of incoherence} $S_\sigma$ is defined as
\begin{equation}
	S_\sigma = 1 - \frac{1}{M}\sum_{m=1}^{M}\bar{s}_m, \qquad  \bar{s}_m = \Theta(\delta - \hat{\sigma}_m),
	\label{eq:Ssig}
\end{equation}
where $\delta = 0.05$ is used throughout this study.  
For the splay, splay-cluster, splay-chimera, and incoherent states, $S_\sigma = 1$, while for the two-cluster, solitary, multi-antipodal, and chimera states, $0 < S_\sigma < 1$. 
The characteristic ranges of the two incoherence measures, $S$ and $S_{\sigma}$, associated with each dynamical state are summarized in Table~\ref{tab1}.

\begin{table}[!ht]
	\centering
	\renewcommand{\arraystretch}{1.7}
	\setlength{\tabcolsep}{6pt}
	\begin{tabular}{ccc}
		\hline\hline
	    Dynamical state & $S$ & $S_\sigma$ \\ \hline
		Two-cluster & $S = 0$ & $0 < S_\sigma < 1$\\
		Solitary & $0 < S \ll 1$ & $0 < S_\sigma < 1$  \\
		Multi-antipodal & $0 < S \ll 1$ & $0 < S_\sigma < 1$ \\
		Chimera & $0 < S < 1$ & $0 < S_\sigma < 1$ \\
		Splay & $S = 0$ & $S_\sigma = 1$  \\
		Splay-cluster & $0 < S \ll 1$ & $S_\sigma = 1$ \\
		Splay-chimera & $0 < S < 1$ & $S_\sigma = 1$ \\
		Incoherent & $S = 1$ & $S_\sigma = 1$ \\ \hline\hline
	\end{tabular}
	\caption{Characterization of the observed dynamical states using the two incoherence measures ($S$, and $S_\sigma$.)}
	\label{tab1}
\end{table}

We observe a dynamical transition from the TC state to the CHI state through the SS and MAC states as a function of $\alpha$ within the range $\beta \in (-0.73\pi, -0.48\pi)$.  
Additionally, a transition sequence TC $\rightarrow$ SS $\rightarrow$ SP is found for $\beta \in (-0.48\pi, -0.26\pi)$.  
In the narrower range $\beta \in (-0.26\pi, 0.0)$, the system transitions from TC to SP via the SPC state.  
For $\beta \in (0.0, 0.2\pi)$, a transition from SPCHI to SP occurs through the SPC state, and for $\beta \in (0.2\pi, 0.35\pi)$, a transition from SPCHI to SPC is observed.  
Beyond $\beta = 0.35\pi$, the INC state dominates for all values of $\alpha$, indicating a complete loss of coherence.

Although it is not feasible to analytically derive stability conditions for all the observed dynamical states, the stability boundary for the TC state can be obtained. By performing a linear stability analysis of Eq. (\ref{sys2}) with $\nu = 1$, we determine the parameter region in which the TC state remains stable. The analysis reveals that, for an $N$-oscillator system with reciprocal coupling, the plasticity dynamics support stable two-cluster solutions within the range $\beta \in (-\pi, 0)$, as detailed in \ref{ap}.

\begin{figure}[!ht]
	\centering
	\includegraphics[width=1.0\linewidth]{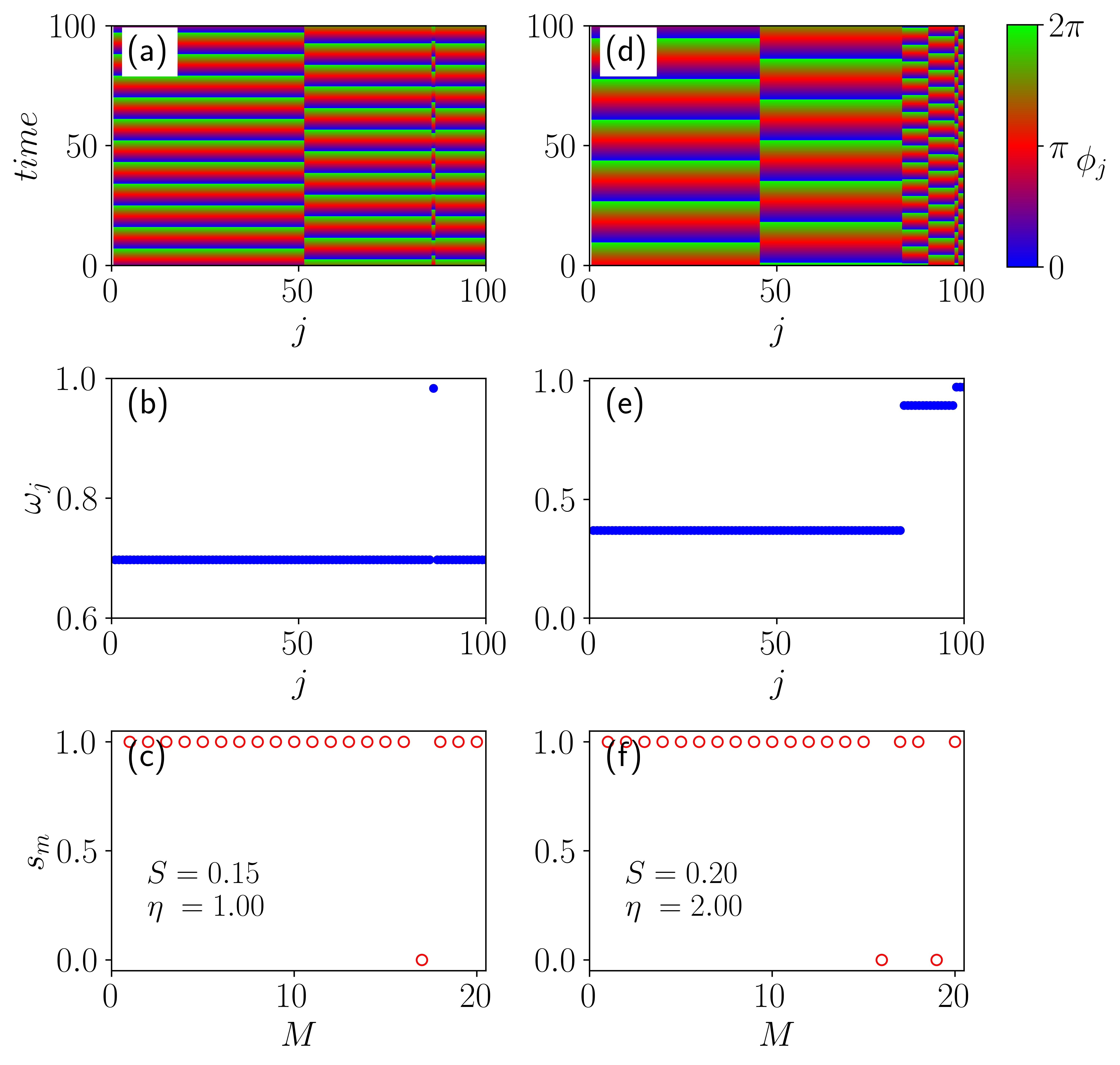}
	\caption{First row: space-time evolution of the phases. Second row: averaged frequency. Third row: distribution of $s_m$ values corresponding to the frequencies. Panels (a)--(c) depict the SS state, while panels (d)--(f) depict the MAC state..}
	\label{Fig10}
\end{figure}

To further distinguish dynamical states that exhibit similar values of the incoherence measures $S$ and $S_{\sigma}$, particularly the SS and MAC states, we incorporate the discontinuity measure (DM) $\eta$, originally introduced in Ref. \cite{gopal2014observation}. While both SS and MAC states satisfy $0 < S \ll 1$ and $0 < S_{\sigma} < 1$, their internal coherence structures differ fundamentally. The DM provides a quantitative indicator of the number of transitions between coherent and incoherent bins and is defined as
\begin{equation}
	\eta = \frac{1}{2}\sum_{i=1}^{M}\left| s_{i+1} - s_i \right|, \qquad s_{M+1} = s_1.
	\label{DM}
\end{equation}
A solitary state contains exactly one bin in which the solitary oscillator is located, producing a single discontinuity between coherent and incoherent bins. Consequently, the DM takes the value $\eta = 1$.
In contrast, a MAC consists of several coherent groups separated by phase gaps that produce multiple discontinuities between coherent and incoherent bins. Therefore, the DM satisfies $\eta > 1$. The $s_m$ distributions for the SS and MAC states are shown in Fig.~\ref{Fig10}.

In Fig.~\ref{Fig10}, panels (a) and (d) display the spatiotemporal patterns for the SS and MAC states, respectively, while panels (b) and (e) show their averaged frequency profiles.  
The corresponding $s_m$ distributions are presented in panels (c) and (f).  
Throughout this work, we consider a system of $N = 100$ oscillators, which are divided into $M = 20$ bins, each containing $n = 5$ oscillators. The standard deviation $\sigma_m$ for each bin is computed using Eq.~\ref{eq:sig_m}, and the $s_m$ values for each bin are shown in the third row of Fig.~\ref{Fig10}. 

Thus, the DM provides an essential refinement for classifying states that appear similar under conventional incoherence measures but differ qualitatively in their spatial organization.

\begin{figure*}[!ht]
	\centering
	\includegraphics[width=1.0\linewidth]{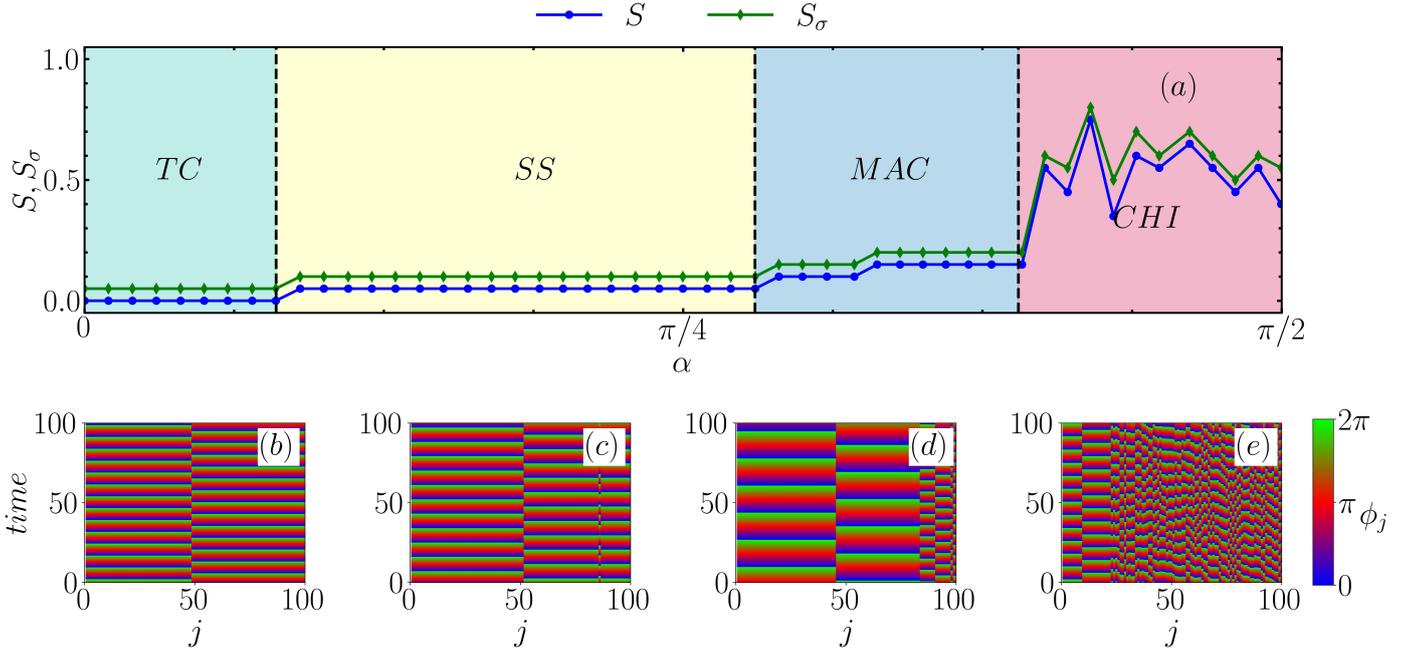}
	\caption{Dynamical transitions as a function of the phase lag $\alpha$. (a) Two incoherence measures ($S$, and $S_\sigma$) plotted against $\alpha$. (b)--(e) Space-time plots of distinct dynamical states for different values of $\alpha$ under the Hebbian rule ($\beta = -0.5\pi$).}
	\label{1p_1}
\end{figure*}

\begin{figure*}[!ht]
	\centering
	\includegraphics[width=1.0\linewidth]{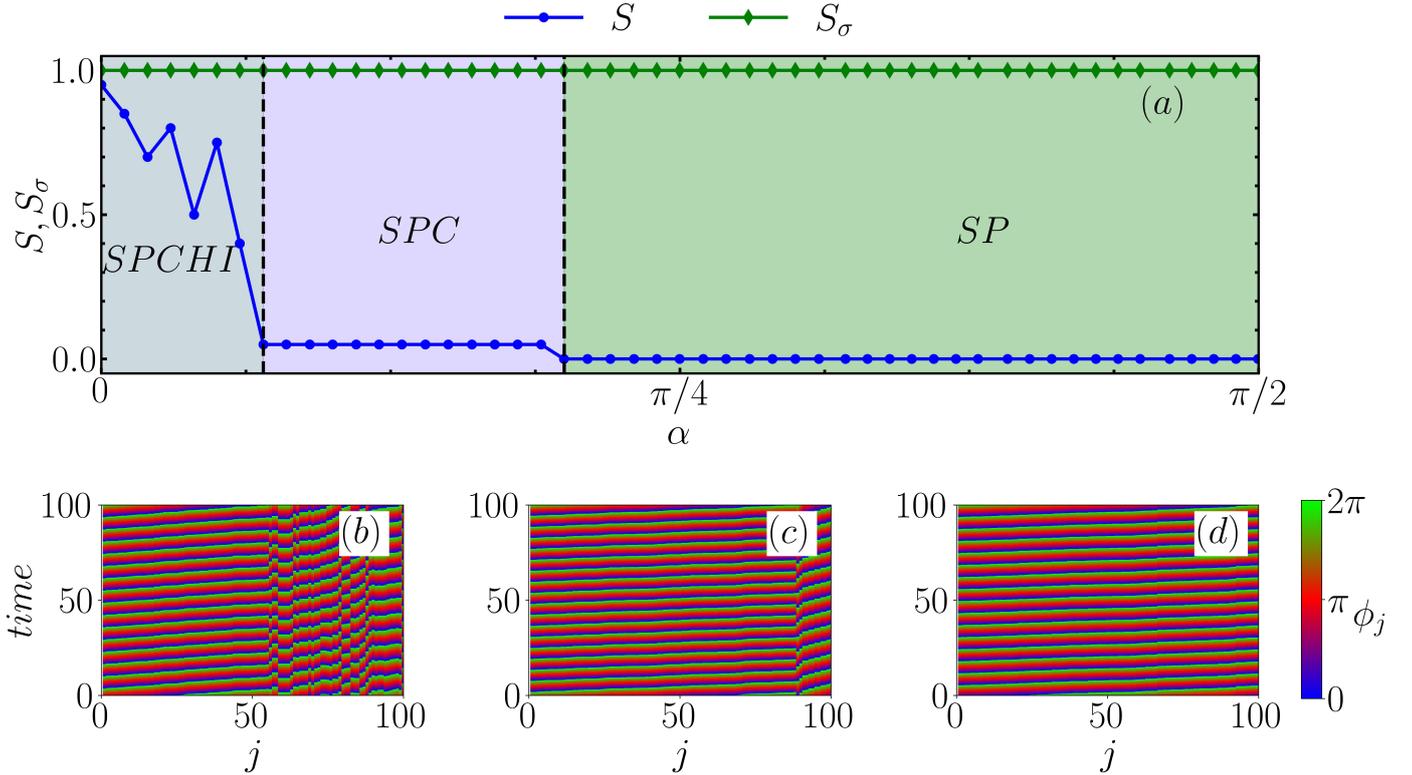}
	\caption{Dynamical transitions as a function of the phase lag $\alpha$. (a) Two incoherence measures ($S$, and $S_\sigma$) plotted versus $\alpha$. (b)--(e) Space--time plots of distinct dynamical states for varying $\alpha$ under the STDP rule ($\beta = 0$).}
	\label{1p2}
\end{figure*}

Following the global mapping of dynamical states in the two-parameter phase diagram, we further examine the dynamical transitions governed by the Hebbian and spike-timing-dependent plasticity (STDP) rules as a function of the phase-lag parameter $\alpha$.  
These transitions are presented in one-parameter diagrams in Figs.~\ref{1p_1} and \ref{1p2} for the Hebbian and STDP cases, respectively, with $\nu = 1.0$ and $\epsilon = 0.01$.

In Fig.~\ref{1p_1}(a), the variations of $S$, and $S_\sigma$ with $\alpha$ confirm the sequence of transitions illustrated in the spatiotemporal plots [Figs.~\ref{1p_1}(b)--\ref{1p_1}(e)].  
The TC state appears for $\alpha \in (0.0, 0.08\pi)$, followed by a transition to the SS state in $\alpha \in (0.08\pi, 0.28\pi)$.  
As $\alpha$ increases further, the system exhibits the MAC state for $\alpha \in (0.28\pi, 0.39\pi)$ and the CHI state for $\alpha \in (0.39\pi, 0.5\pi)$.

For $\beta = 0$ [Fig.~\ref{1p2}(a)], corresponding to the STDP rule, the SPCHI state is observed for $\alpha \in (0.0, 0.07\pi)$, transitioning to the SPC state via the SP state as $\alpha$ increases.  
The SPC state remains stable in $\alpha \in (0.07\pi, 0.2\pi)$, while the SP state persists for $\alpha \in (0.2\pi, 0.5\pi)$.  
The corresponding spatiotemporal patterns for the SPCHI, SPC, and SP states are shown in Figs.~\ref{1p2}(b)--\ref{1p2}(d).  
Beyond $\beta = 0.5\pi$, where the system follows the anti-Hebbian rule, only the incoherent (INC) state is observed.  
At $\beta = 0.5\pi$, the INC state appears for all $\alpha$, as illustrated in Fig.~\ref{Fig8} for $\alpha = 0.4\pi$ and $\beta = 0.5\pi$.  
These dynamical transitions correspond to the red and blue dashed trajectories shown in Fig.~\ref{2p}.

\section{Conclusion}
\label{conclusion}

In this work, we have systematically investigated the collective dynamics of a system of identical pendulum oscillators with adaptive coupling, where the coupling strengths coevolve with the oscillator states according to Hebbian and STDP learning rules. The inclusion of a phase-lag parameter $\alpha$ enables the emergence of diverse synchronization patterns and rich dynamical transitions. The adaptive network exhibits a variety of collective dynamical states, including the two-cluster, solitary, multi-antipodal, chimera, splay, splay-cluster, splay-chimera, and incoherent states. These states were analyzed through spatiotemporal plots, averaged frequency profiles, phase snapshots, and coupling matrix representations, revealing the intricate interplay between phase lag and plastic adaptation.

By constructing two-parameter phase diagrams in the $(\alpha, \beta)$ space, we demonstrated how different adaptive mechanisms yield distinct dynamical regimes. Specifically, we identified several transition sequences such as 
TC $\rightarrow$ SS $\rightarrow$ MAC $\rightarrow$ CHI, TC $\rightarrow$ SS $\rightarrow$ SP,  TC $\rightarrow$ SPC $\rightarrow$ SP,  SPCHI $\rightarrow$ SPC $\rightarrow$ SP,  and SPCHI $\rightarrow$ SPC, depending on the adaptation control parameter $\beta$ and the corresponding ranges of $\alpha$.  To characterize these transitions, we employed two complementary incoherence measures, the frequency-based metric $S$ and the instantaneous-phase-based metric $S_{\sigma}$, which together enable a comprehensive and systematic classification of the collective states. Additionally, to distinguish between closely related states, such as the solitary and multi-antipodal configurations, we introduced a discontinuity measure based on the distribution of local coherence indicators.

A significant and novel finding of this study is the emergence of the solitary state in adaptive oscillator networks \textit{without} the inclusion of time-delay, parameter mismatch, or noise, features that are typically required for such states in previous works.  Earlier studies on adaptive or plastic oscillator networks have primarily focused on chimera or cluster synchronization patterns under Hebbian or anti-Hebbian plasticity mechanisms. In contrast, our results reveal that even in the absence of heterogeneity or delay, the interplay between the phase lag and the plastic adaptation rules alone is sufficient to induce solitary and multi-antipodal states, thereby enriching the known landscape of adaptive network dynamics.

We also analytically derived the stability condition for the two-cluster state, confirming that it remains stable within the range $\beta \in (-\pi, 0)$ for all $\alpha$. This analytical boundary aligns with the numerically observed transition region in the phase diagram, providing a theoretical foundation for the stability of frequency-entrained states in adaptive pendulum networks.

The observed coexistence of multiple attractors under identical system parameters highlights the phenomenon of extreme multistability, implying that adaptive plasticity can serve as a mechanism for memory storage and switching among collective states. Such features could be relevant for understanding information encoding in adaptive neuronal or mechanical networks.

Future research directions include extending this framework to multilayer or modular adaptive networks to explore inter-layer synchronization and competition between adaptation rules. In addition, incorporating explicit restoring forces, time delays, or stochastic fluctuations may shed light on how robustness and adaptability coexist in complex adaptive systems. Further analytical studies could focus on bifurcation structures underlying the observed transitions and on identifying universal scaling behaviors that govern the onset of multistability and chimera formation.

Overall, the present study broadens the understanding of how adaptive plasticity mechanisms shape collective dynamics, revealing that even simple pendulum oscillators with evolving couplings can produce a rich variety of coherent, incoherent, and hybrid dynamical states.
	
\section*{Acknowledgments}
R. A. acknowledges SASTRA for providing Teaching Assistantship. The research contributions of R. S. and V.K.C. are part of a project funded by the SERB-CRG (Grant No. CRG/2022/004784). The authors gratefully acknowledge the Department of Science and Technology (DST), New Delhi, for providing computational facilities through the DST-FIST program under project number SR/FST/PS-1/2020/135, awarded to the Department of Physics.
	
\section*{Data Availability}
	The data that support the findings of this study are available from the corresponding author upon reasonable request.

\appendix
\section{Extreme Multistability and Stability Condition}
\label{ap}
\begin{figure}[!ht]
	\centering
	\includegraphics[width=1.0\linewidth]{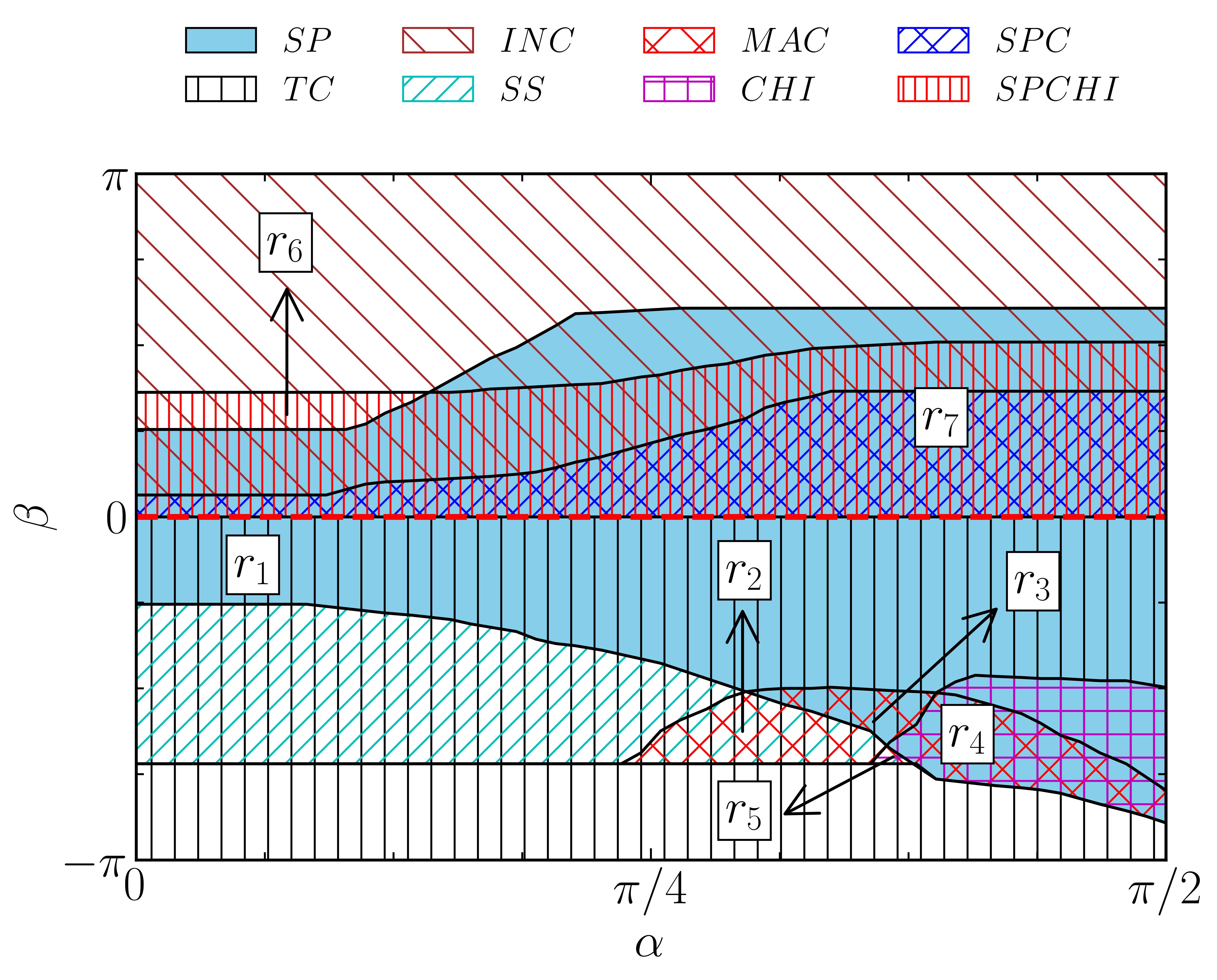}
	\caption{Two-parameter phase diagram of dynamical states in the $(\alpha, \beta)$ space, obtained from ten realizations of random initial conditions. The initial conditions are identical to those used in Fig.~\ref{2p}, but multistability emerges among the observed states depending on the specific choice of initial conditions. All other parameters are identical to those used in Fig.~\ref{2p}.}
	\label{2p-multi}
\end{figure}

Adaptive networks exhibit pronounced multistability among various collective dynamical states, strongly depending on the initial conditions. This property is illustrated in the two-parameter phase diagram shown in Fig.~\ref{2p-multi}, plotted in the $(\alpha, \beta)$ parameter space.

Figure~\ref{2p-multi} depicts the coexistence of several distinct dynamical states across ten realizations of random initial conditions.  
The multistable regions are defined as follows:  $r_1$ represents multistability between the TC and SP states, observed for $\beta \in (-0.75\pi, 0.0)$ across all $\alpha$ values;  $r_2$ indicates coexistence among TC, SS, and MAC states;  $r_3$ denotes multistability among TC, SP, and MAC states;  $r_4$ corresponds to the coexistence of TC, SP, MAC, and CHI states;  $r_5$ shows coexistence among TC, SS, MAC, and CHI states;  $r_6$ reflects multistability between SPCHI and INC states;  and $r_7$ denotes coexistence among SP, SPC, SPCHI, and INC states. Additionally, coexistence between TC and SS states is observed for $\beta \in (-0.75\pi, -0.26\pi)$ and $\alpha \in (0.0, 0.23\pi)$. Overall, Fig.~\ref{2p-multi} clearly illustrates the diversity of multistable regions obtained under random initial conditions.

Although determining stability boundaries for all observed dynamical states is not feasible, the stability condition of the TC state can be derived analytically.  
Starting from Eq.~(\ref{sys2}) with $\nu = 1.0$, the governing equations are written as
\begin{align}
	\ddot{\phi}_i + \dot{\phi}_i &= 1 - \frac{1}{N}\sum_{j=1}^{N} k_{ij}\sin(\phi_i - \phi_j + \alpha), \nonumber \\
	\dot{k}_{ij} &= -\epsilon\,[\sin(\phi_i - \phi_j + \beta) + k_{ij}], \qquad |k_{ij}| \le 1.
	\label{sys3}
\end{align}

To analyze stability, Eq.~(\ref{sys3}) is transformed into a first-order system by introducing $\dot{\phi}_i = \omega_i$.  
The resulting equations are
\begin{align}
	\dot{\phi}_i &= \omega_i, \nonumber \\
	\dot{\omega}_i &= -\omega_i + 1 - \frac{1}{N}\sum_{j=1}^{N} k_{ij}\sin(\phi_i - \phi_j + \alpha).
	\label{sys4}
\end{align}

Linearizing around the steady-state configuration yields the Jacobian matrix.  
The diagonal element of the Jacobian corresponding to oscillator $i$ is expressed as
\begin{equation}
	DF_{ii} = -\frac{1}{N}\sum_{j \neq i} k_{ij}^*\cos(\phi_i - \phi_j + \alpha),
	\label{jacob}
\end{equation}
where the superscript $*$ denotes quantities evaluated at steady state.

Assume that the network splits into two clusters:  
cluster 1 containing $N_1$ oscillators and cluster 2 containing $N_2 = N - N_1$ oscillators.  
The phases of the two clusters are denoted by $\phi_A$ and $\phi_B = \phi_A + \pi$, respectively, and the two clusters rotate with a common frequency $\omega$ in the steady state.  
For the TC (two-cluster) state: for oscillators $i$ and $j$ in the same cluster, the steady coupling strength and phase difference are $k_{ij}^* = -\sin\beta$ and $\phi_{ij}^* = 0$;  for oscillators in opposite clusters, $k_{ij}^* = \sin\beta$ and $\phi_{ij}^* = \pi$.

Under these assumptions, the Jacobian matrix elements are given by
\begin{equation}
	DF_{ij} =
	\begin{cases}
		\dfrac{N-1}{N}\sin\beta\cos\alpha, & \text{if } i = j, \\[6pt]
		\dfrac{1}{N}\sin\beta\cos\alpha, & \text{if } i \ne j.
	\end{cases}
	\label{DF}
\end{equation}

The corresponding eigenvalues of the Jacobian are
\begin{equation}
	\lambda_k =
	\begin{cases}
		0, & k = 0, \\[6pt]
		\dfrac{-1 + \sqrt{1 + 4\sin\beta\cos\alpha}}{2}, & k \ne 0.
	\end{cases}
	\label{eig}
\end{equation}

For $\lambda_k = 0$, the marginal stability condition of the TC state occurs at $\beta = 0$.  
Thus, the TC state remains stable for $\beta \in (-\pi, 0)$ for all $\alpha$.  

This analytical stability boundary corresponds to the red dashed line shown in the phase diagram in Fig.~\ref{2p-multi}.

	\bibliographystyle{elsarticle-num} 
	\bibliography{ref}
	
\end{document}